\documentclass[pra,twocolumn,superscriptaddress]{revtex4}

\usepackage{amsmath}
\usepackage{latexsym}
\usepackage{amssymb}
\usepackage{pdfpages}
\usepackage{graphicx}
\usepackage{subfigure}
\usepackage{epstopdf}
\usepackage{pgffor}
\usepackage{amsfonts}
\usepackage{textcomp}
\usepackage{comment}
\usepackage{braket}
\usepackage{booktabs}
\usepackage{bbm}
\usepackage{bm}
\usepackage{xcolor}
\definecolor{myurlcolor}{rgb}{0,0,0.8}
\definecolor{mycitecolor}{rgb}{0,0,0.8}
\definecolor{myrefcolor}{rgb}{0,0,0.8}
\usepackage{hyperref}
\usepackage[capitalise]{cleveref}
\hypersetup{colorlinks,
linkcolor=myrefcolor,
citecolor=mycitecolor,
urlcolor=myurlcolor}

\usepackage{supertabular}
\usepackage[draft]{fixme}
\usepackage{amsmath,bbm}%,mathrsfs}
\usepackage{graphicx}
\usepackage{amsfonts}
\usepackage{amssymb}
\usepackage{amsmath, amssymb, amsthm,verbatim,graphicx,bbm}
\usepackage{mathrsfs}
\usepackage{color,xcolor,longtable}

\makeatletter
\usepackage{pifont}
\makeatother

\usepackage{epstopdf}
\usepackage{subfigure}
\usepackage{appendix}

\makeatletter
\usepackage{pifont}
\makeatother

\usepackage{dcolumn}
\usepackage{epstopdf}

\begin{document}

\title{Towards Robust Optimal Measurements Against Noise in Quantum Metrology}

\author{Xinglei Yu}
\affiliation{School of Physical Science and Technology, Ningbo University, Ningbo, 315211, China}
\affiliation{Department of Mechanical and Automation Engineering, The Chinese University of Hong Kong, Shatin, Hong Kong, 999077, China}
\author{Xinzhi Zhao}
\affiliation{School of Physical Science and Technology, Ningbo University, Ningbo, 315211, China}
\author{Liangsheng Li}
\email{liliangshengbititp@163.com}
\affiliation{National Key Laboratory of Scattering and Radiation, Beijing 100854, China}
\author{Stanis{\l}aw Kurdzia{\l}ek}
\affiliation{Faculty of Physics, University of Warsaw, Pasteura 5, 02-093 Warszawa, Poland}
\author{Chengjie Zhang}
\email{cjzhang@ustc.edu}
\affiliation{School of Physical Science and Technology, Ningbo University, Ningbo, 315211, China}
\affiliation{Hefei National Laboratory, University of Science and Technology of China, Hefei 230088, China}
\author{Chuan-Feng Li}
\email{cfli@ustc.edu.cn}
\affiliation{CAS Key Laboratory of Quantum Information, University of Science and Technology of China, Hefei 230026, China}
\affiliation{CAS Center for Excellence in Quantum Information and Quantum Physics University of Science and Technology of China, Hefei 230026, China}
\affiliation{Hefei National Laboratory, University of Science and Technology of China, Hefei 230088, China}
\author{Guang-Can Guo}
\affiliation{CAS Key Laboratory of Quantum Information, University of Science and Technology of China, Hefei 230026, China}
\affiliation{CAS Center for Excellence in Quantum Information and Quantum Physics University of Science and Technology of China, Hefei 230026, China}
\affiliation{Hefei National Laboratory, University of Science and Technology of China, Hefei 230088, China}

\begin{abstract}
Quantum parameter estimation utilizes quantum mechanical effects to attain higher measurement precision than classical schemes. In practical implementations, however, noise is inevitably present during the measurement process, causing a decrease in precision. Quantifying the impact of noise on different measurements is of considerable significance. Here, we experimentally investigate robust optimal measurements based on the theory of Fisher information measurement noise susceptibility (FI MENOS), which quantifies how susceptible a measurement is to noise. By constructing a polarizing Mach-Zehnder interferometer, we implement phase estimation under controlled noise. Our results indicate that different measurements exhibit distinct sensitivities to noise. To assess the influence of diverse noise types on precision, we further construct an experimental setup capable of introducing various forms of noise. The experimental results affirm that FI MENOS represents the worst-case scenario for estimation precision, enabling us to evaluate the noise immunity of optimal measurements. Our work provides a deeper insight into quantum metrology with noise, marking a notable advancement in quantifying the robustness of quantum estimation schemes against measurement noise effects.
\end{abstract}

\maketitle

\section{Introduction}\label{section1}
As the core driving force behind the next-generation information revolution, quantum technologies have achieved landmark advances in quantum computing \cite{computing1,computing2,computing3}, the fundamental verification of quantum mechanics \cite{wave1,wave2}, quantum communication \cite{communication1,communication2}, and quantum metrology. These technologies exhibit far-reaching applications in both frontier physics and industrial practice. Among these quantum technologies, quantum metrology is a significant quantum technology in modern science and engineering, which aims to improve the measurement precision. It has been applied in various fields, including atomic masses \cite{atmm}, the fine-structure constant $\alpha$ \cite{fstruc1,fstruc2}, searches for dark matter and dark energy \cite{dakm1,dakm2}, and gravitational-wave detection \cite{gw1,gw2,gw3,gw4}, among others. Quantum parameter estimation represents a crucial area of research within quantum metrology, where the ultimate precision is limited by quantum Cram\'{e}r-Rao bound (QCRB), an extension of the classical Cram\'{e}r-Rao bound (CRB) in classical parameter estimation \cite{qem1,qem2,qem3,qem4}. Based on this foundation, researchers have investigated and achieved the Heisenberg limit using quantum resources, which surpasses the precision limit of classical schemes \cite{HL1,HL2,HL3,HL4,entgHL1,entgHL2,seqHL1,seqHL2,seqHL3,seqHL4,seqHL5}. A new quantum resource known as indefinite causal order has recently been proposed, enabling the possibility of even reaching the super-Heisenberg limit for infinite dimensional Hilbert spaces \cite{supHL1,supHL2,supHL3}. Additionally, non-Hermitian quantum metrology, an emerging field with the potential to revolutionize precision measurement, has also been extensively studied in recent years \cite{nqs1,nqs2,nqs3,nqs4,nqs5,nqs6,nhqsXinglei}. While these advancements are remarkable, it is essential to acknowledge that noise inevitably undermines precision in quantum estimation.

The QCRB is always attainable by performing optimal measurements. However, the realization of these measurements is typically imperfect, thereby impacting the ultimate precision. In previous studies, noise has been classified into different types, e.g, detector dark counts, measurement output crosstalks, classical readout noise, etc \cite{noise1,sensor}. Particularly, the effect of readout noise in quantum estimation was systematically studied in Ref.~\cite{noise2}. Nevertheless, there remains a necessity to investigate the impact of noise in a more general manner. Recently, researchers have theoretically conducted a generalized and fundamental investigation of measurement noise without assuming its specific form, and proposed a quantity called Fisher information measurement noise susceptibility (FI MENOS) \cite{noise3}. It provides a way to assess the robustness of quantum metrology scheme. To mitigate dissipative noise, recent studies have investigated non-Markovian effects in Mach-Zehnder interferometers \cite{an1} and Floquet engineering schemes \cite{an2}.  Noise has been experimentally investigated in various quantum technology fields, including quantum computation \cite{qcom1,qcom2,qcom3,qcom4}, quantum communication \cite{qcommu1,qcommu2,qcommu3,qcommu4,qcommu5} and quantum control \cite{qcontr1,qcontr2,qcontr3,qcontr4}. However, the experimental research on noise in quantum metrology is still very limited \cite{noiseqm1,noiseqm2}, particularly regarding measurement noise.

In this work, we experimentally investigate robust optimal measurements based on FI MENOS theory. We implement phase estimation in polarizing Mach-Zehnder interferometer under different intensity of noise which is effectively controlled by the interference visibility. Repeating the estimation process for different measurements, we observe that the measurements corresponding to the lower values of FI MENOS are less susceptible to noise. Furthermore, we also compare the estimation precision under different forms of noise, the experimental results show that FI MENOS represents the worst-case scenario. The measurement exhibiting the highest noise immunity can be determined. Our work advances the study of quantum metrology in the presence of noise.

\section{Fisher information measurement noise susceptibility}\label{sec:3}
For a final state $\rho_\theta$ encoded with unknown parameter $\theta$, the estimation precision is limited by CRB  $(\Delta\hat{\theta})^2\geq1/(\nu F_C)$, where $\hat{\theta}$ is an unbiased estimator of $\theta$, $\nu$ is the number of measurements, and $F_C$ is the Fisher information (FI). Obviously, FI is an important quantity to characterize the precision, it can be calculated by
\begin{eqnarray}
F_C=\sum\limits_{i}p_il_i^2,\quad p_i=\mathrm{Tr}[\hat{M}_i\rho_\theta],\label{expesFI}
\end{eqnarray}
where $\hat{M}_i$ is the measurement operator we performed and $l_i=\mathrm{Tr}[\dot{\rho}_\theta\hat{M}_i]/\mathrm{Tr}[\rho_\theta\hat{M}_i]$. The quantum measurements can be characterized by positive operator-valued measurements (POVMs) $\bm{M}=\{\hat{M}_i|\hat{M}_i\geq0, \sum_i\hat{M}_i=I\}$, where $i$ denotes the $i$th measurement outcome. In quantum systems, we can always find some special POVMs to maximize the FI, then we obtain the QCRB $(\Delta\hat{\theta})^2\geq1/(\nu F_Q)$ \cite{qem2,qem3,qem4}, where $F_Q$ is quantum Fisher information (QFI), these special POVMs are called optimal measurements. For a fixed probe state and quantum system, the FI only depends on the measurement, here we denote the FI that corresponds to the specific measurement $\bm{M}$ as $F_C[\bm{M}]$.

\begin{figure}[t]
  \centering
  \includegraphics[width=0.5\textwidth]{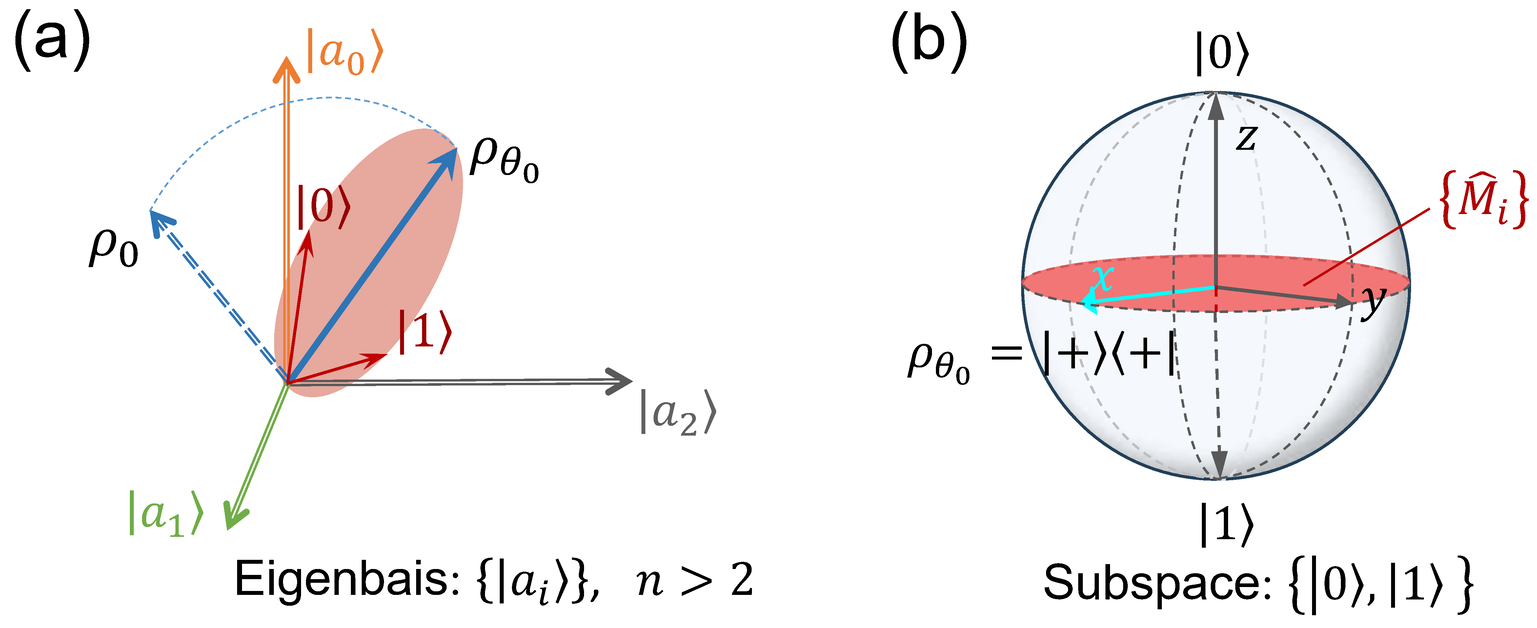}
  \caption{Schematic illustration of the pure state model. (a) The decomposition of final state. For a fixed value of a parameter $\theta=\theta_0$, the final state $\rho_{\theta_0}$ can be decomposed into a two level subspace consists of $\ket{0}=|\psi_{\theta_0}\rangle/\sqrt{2}+i{\scriptstyle\sqrt{2F_Q^{-1}}}(\ket{\dot{\psi}_{\theta_0}}-\langle\psi_{\theta_0}|\dot{\psi}_{\theta_0}\rangle|\psi_{\theta_0}\rangle)$ and $\ket{1}=|\psi_{\theta_0}\rangle/\sqrt{2}-i{\scriptstyle\sqrt{2F_Q^{-1}}}(\ket{\dot{\psi}_{\theta_0}}-\langle\psi_{\theta_0}|\dot{\psi}_{\theta_0}\rangle|\psi_{\theta_0}\rangle)$. (b) The Bloch sphere in the basis of $\ket{0}$ and $\ket{1}$. On this Bloch sphere, the final state $\rho_{\theta_0}$ can be written as $|+\rangle\langle+|$, and the set of optimal measurements is the equator.}\label{ps}
\end{figure}

To saturate QCRB, finding the optimal measurement is inevitable. In most cases, it is not difficult to find optimal measurements. For a final state $\rho_\theta$, an optimal measurement is performed when we measure in the eigenbasis of the symmetric logarithmic derivative (SLD) $\hat{L}_\theta$, where SLD is the operator that satisfies the equation $(\hat{L}_{\theta}\rho_{\theta}+\rho_{\theta}\hat{L}_{\theta})/2=\partial_{\theta}\rho_{\theta}$ \cite{qem3,qem4}. It is worth noticing that this equation could have many solutions, which means the optimal measurement is not unique in general.

\begin{figure*}[t]
  \centering
  \includegraphics[width=1\textwidth]{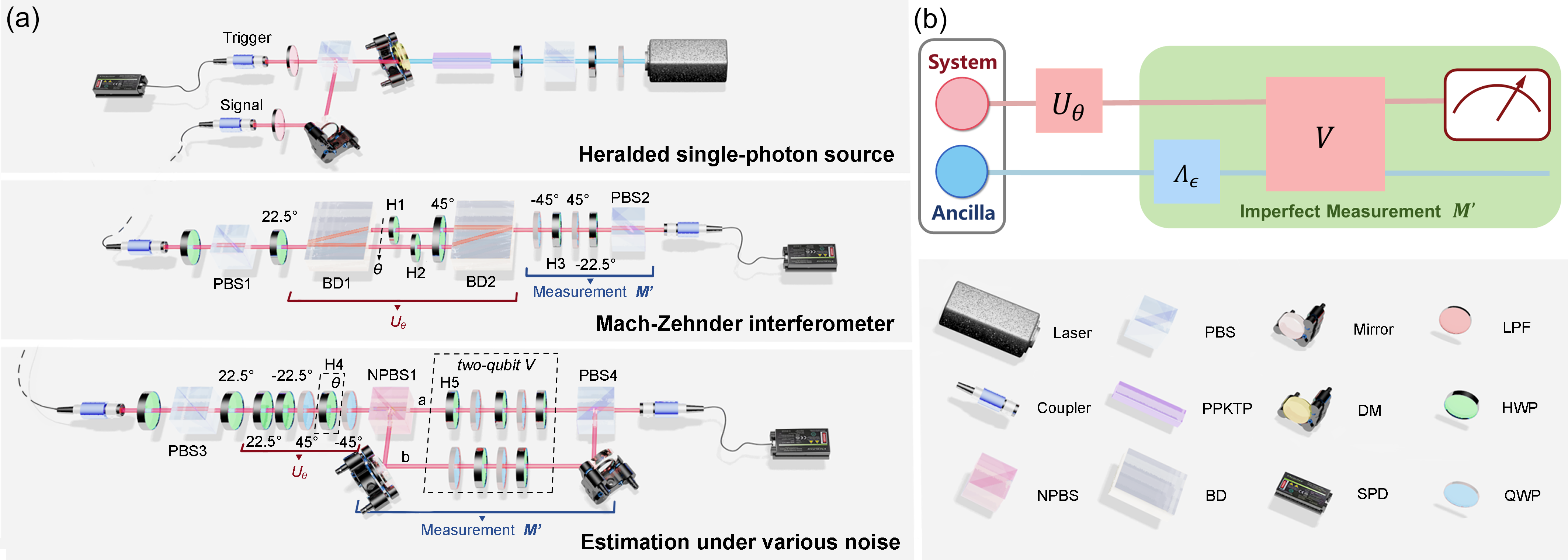}
  \caption{(a) Experimental setup. The top layer presents the implementation of heralded single-photon source, photon pairs were generated via spontaneous parametric down-conversion. The middle layer presents the implementation of polarizing Mach-Zehnder interferometer. The polarizing Mach-Zehnder interferometer is conducted with two beam displacers (BDs), so signal photons should be purified by half-wave plates (HWPs) and  polarization beam splitter1 (PBS1) firstly, and then prepared as $(\ket{0}+\ket{1})/\sqrt{2}$ by HWP ($22.5^\circ$). After BD1, the photon is separated into two paths. The relative phase is introduced by the optical path difference between two paths (including the thickness difference between H1 and H2). The bottom layer is the experimental setup which is able to change the form of noise. The relative phase is controlled by the angle of H4. Using noisy channel and two-qubit operator $V$, we equivalently realize the imperfect measurement $\bm{\tilde{M}}$. (b) The quantum circuit of realizing the imperfect measurement. Dichroic mirror (DM); single photon detector (SPD); long-wave path filter (LPF).}\label{exp}
\end{figure*}

However, in practical experiments, quantum measurements $\bm{M}$ cannot be perfectly implemented, which leads to our inability to achieve CRB. The imperfect measurement can be written as $\bm{\tilde{M}}=(1-\epsilon)\bm{M}+\epsilon\bm{N}$, which means the undesired measurements $\bm{N}$ will be performed with a small probability $\epsilon\ll1$. This formula could describe various types of noise, such as random rotations of measurement basis, dark counts and signal losses \cite{noise3}. To quantify the effect of noise $\bm{N}$, we can use
\begin{eqnarray}
\chi[\bm{M},\bm{N}]=\lim\limits_{\epsilon\to 0}\frac{F_C[\bm{M}]-F_C[(1-\epsilon)\bm{M}+\epsilon\bm{N}]}{\epsilon F_C[\bm{M}]},\label{chiMN}
\end{eqnarray}
which can be understood as the relative decrease of FI under the effect of infinitesimally noise $\bm{N}$ \cite{noise3}. This quantity depends on the specific unexpected measurements $\bm{N}$, which implies that the quantum measurements $\bm{M}$ exhibit varying susceptibilities to diverse forms of noise. As we discussed above, the optimal measurement is not unique. To compare the noise resisting ability of different optimal measurements without considering a specific noise model, we consider the worst-case scenario, i.e., the maximal shrinkage rate of the FI caused by an infinitesimal measurement noise described by arbitrary POVMs $\bm{N}$. The answer is given by a quantity called Fisher information measurement noise susceptibility (FI MENOS) which is proposed in Ref.\;\cite{noise3}, its definition is
\begin{eqnarray}
\chi[\bm{M}]=\max\limits_{\bm{N}\in\mathcal{M}}\chi[\bm{M},\bm{N}],\label{FIMENOS}
\end{eqnarray}
where $\mathcal{M}$ represents the set of all POVMs. The explicit formula of FI MENOS is $\chi[\bm{M}]=1+(l_1^2+l_K^2+\|A_1-A_K\|_1)/2F_C[\bm{M}]$, where $A_i=l_i^2\rho_\theta-2l_i\dot{\rho}_\theta$, $\|A\|_1=\mathrm{Tr}\sqrt{AA^\dagger}$, $K$ is the number of measurement outcomes and $l_1\leq l_2\leq\ldots\leq l_K$ \cite{noise3}. FI MENOS characterizes the susceptibility of FI to small noise in measurement, the larger $\chi[\bm{M}]$ means that FI is more susceptible to noise. It enables us to assess the noise immunity of measurement. Recently, the FI MENOS theory has been generalized to multi-parameter quantum estimation \cite{multi-FIMENOS}. Note that FI MENOS is defined in the limit of $\epsilon\rightarrow0$ and may therefore be ineffective under strong noise, as the second-order term $o(\epsilon^2)$ can no longer be neglected.

For pure state models, one can always construct two orthonormal vectors $\ket{0}$, $\ket{1}$ span ($\ket{\psi_{\theta_0}}$, $\ket{\dot{\psi}_{\theta_0}}$) even though the dimension of Hilbert space $n>2$, where $\theta_0$ is the known parameter value that is close to $\theta$, as shown in Figure\;\ref{ps} (a) and (b). Then, we have $\rho_{\theta_0}=|+\rangle\langle+|$, $\dot{\rho}_{\theta_0}=\sqrt{F_Q}\sigma_y/2$, and the optimal measurements can be written as $\hat{M}(\varphi)=|\phi(\varphi)\rangle\langle\phi(\varphi)|$, where $\ket{+}=(\ket{0}+\ket{1})/\sqrt{2}$, $\sigma_y$ is a Pauli matrix and $\ket{\phi(\varphi)}=(\ket{0}+e^{i\varphi}\ket{1})/\sqrt{2}$ \cite{optm4}. Thus, we can treat pure state models as a phase estimation in Mach-Zehnder interferometer. It has been theoretically proved that FI MENOS $\chi[\bm{M}]\geq4$ in this case \cite{noise3}. And FI MENOS indicates that for different $\varphi$, the optimal measurements exhibit different susceptibility to noise, and the optimal measurements are robust near $\varphi=\pi/2$. Here, we experimentally implement the phase estimation in polarizing Mach-Zehnder interferometer and demonstrate this conclusion. To illustrate that the performance of noisy parameter estimation would not be worse than the performance given by FI MENOS, we further implement the phase estimation experiment in the experimental setup that allows the change of the form of noise $\bm{N}$.

\section{Experimental setup}\label{sec:4}

Photon pairs (810 nm) were generated through type-II spontaneous parametric down-conversion (SPDC) by pumping a periodically poled $\mathrm{KTiOPO_4}$ (PPKTP) crystal with a 405 nm laser \cite{sps}. One photon serves as the trigger photon, while the other photon is called the signal photon and is directed into the parameter estimation systems, as shown in Figure\;\ref{exp}(a). In our experiment, the system qubit is encoded in the polarization degree of freedom, the horizontally polarized state $\ket{H}$ is encoded as $\ket{0}$, while the vertically polarized state $\ket{V}$ is encoded as $\ket{1}$.

As shown in Figure\;\ref{exp}(a), the middle layer is the experimental setup of polarizing Mach-Zehnder interferometer which is constructed with two BDs, so we need to purify and prepare the initial state as $\ket{\psi_0}=(\ket{0}+\ket{1})/\sqrt{2}$. The relative phase $\theta$ between the upper and lower arm is introduced by the optical path difference, and the final state can be written as $\ket{\psi_\theta}=(\ket{0}+e^{i\theta}\ket{1})/\sqrt{2}$. In this parameterization process, the evolution operator can be written as $\hat{U}_\theta=|0\rangle\langle0|+e^{i\theta}|1\rangle\langle1|$. For optimal measurements $\bm{M}$ which saturate QCRB, the QFI is $F_Q=4(\langle\partial_\theta\psi_\theta|\partial_\theta\psi_\theta\rangle-|\langle\partial_\theta\psi_\theta|\psi_\theta\rangle|^2)=1$. As we mentioned above, the optimal measurements is $\hat{M}(\varphi)=|\phi(\varphi)\rangle\langle\phi(\varphi)|$, which is realized by the combination of HWPs, quarter-wave plates (QWPs) and PBS, where $\varphi$ is well controlled by the angle of H3. By Rotating H3, we perform a set of orthogonal projective measurements $\bm{M}=\{\hat{M}_\pm(\varphi)\}$, where $\hat{M}_\pm(\varphi)=|\phi_\pm(\varphi)\rangle\langle\phi_\pm(\varphi)|$ and $|\phi_\pm(\varphi)\rangle=(\ket{0}\pm e^{i\varphi}\ket{1})/\sqrt{2}$. Notice that our objective is to have measurements influenced by small noise to demonstrate FI MENOS theory. Here, the measurement noise $\epsilon\bm{N}$ is introduced and controlled by the interference visibility $\mathcal{V}$, which is consistent with the theoretical model. The exact form of imperfect measurement we realized is $\bm{\tilde{M}}=\{\hat{M}'_\pm(\varphi)\}$, where $\hat{M}'_\pm(\varphi)=(1-\epsilon)\hat{M}_\pm(\varphi)+\epsilon\hat{M}_\mp(\varphi)$, $\epsilon=(1-\mathcal{V})/2$ and the form of noise $\bm{N}=\{\hat{M}_\mp(\varphi)\}$ is fixed and immutable.

The bottom layer in Figure\;\ref{exp}(a) is the experimental setup capable of introducing various forms of noise $\bm{N}$. The Mach-Zehnder interferometer has been substituted with the combination of HWPs and QWPs, which also realizes the evolution operator $\hat{U}_\theta$, where the parameter $\theta$ is controlled by the angle of H4. In order to implement the imperfect measurement $\bm{\tilde{M}}$, we introduce an ancilla qubit that is encoded in two independent spatial modes of the signal photons $\{|a\rangle,|b\rangle\}$.  The quantum circuit representation of our imperfect measurement realization scheme is shown in Figure\;\ref{exp}(b). In this experiment, the initial state is prepared as $\rho^{SA}_{0}=|\psi_0\rangle\langle\psi_0|\otimes|a\rangle\langle a|$ firstly, after evolution operator $\hat{U}_\theta$, it becomes $\rho^{{SA}}_{\theta}=|\psi_\theta\rangle\langle\psi_\theta|\otimes|a\rangle\langle a|$. The noisy channel on ancilla qubit $\Lambda_\epsilon(\cdot)=\sum_iK_i\cdot K_i^\dagger$ introduce a noise with intensity $\epsilon$, where Kraus operators read $K_1=I\otimes\sqrt{\epsilon} I$, $K_2=I\otimes\sqrt{1-\epsilon}\left(|a\rangle\langle b|+|b\rangle\langle a|\right)$.
Via performing a two-qubit unitary operator $\hat{V}=\hat{U}_1\otimes|a\rangle\langle a|+\hat{U}_2\otimes|b\rangle\langle b|$ and the projective measurement $\hat{P}_0\otimes I$ on this two-qubit state, we equivalently perform the imperfect measurement $\bm{\tilde{M}}$ on system qubit $\mathrm{Tr_S}\{\mathrm{Tr_A}[\hat{V}(\sum_iK_i\rho^{SA}_{\theta}K_i^\dagger)\hat{V}^\dagger(\hat{P}_0\otimes I)]\}=\mathrm{Tr_S}\{|\psi_\theta\rangle\langle\psi_\theta|\bm{\tilde{M}}\}$, where $\hat{P}_0=|0\rangle\langle0|$, $\bm{\tilde{M}}=\mathrm{Tr_A}[\sum_iK_i^\dagger\hat{V}^\dagger(\hat{P}_0\otimes|a\rangle\langle a|) \hat{V}K_i]=(1-\epsilon)\emph{\textbf{M}}+\epsilon\emph{\textbf{N}}$, $\bm{N}=\hat{U}_1\hat{P}_0\hat{U}_1^\dagger$ and $\bm{M}=\hat{U}_2\hat{P}_0\hat{U}_2^\dagger$. We can find that the form of noise depends on the unitary operator $\hat{U}_{1}$, which corresponding to the combination of HWPs and QWPs in path $a$, as shown in the bottom layer of Figure\;\ref{exp}(a). By changing the operator $\hat{U}_1$, we could change the form of noise $\bm{N}$. In our experimental setup, the noisy channel $\Lambda_\epsilon$ is realized by nonpolarizing beam splitter1 (NPBS1).

\section{Phase estimation and FI MENOS in Mach-Zehnder interferometer}\label{sec:5}

\begin{figure*}[t]
  \centering
  \includegraphics[width=1\textwidth]{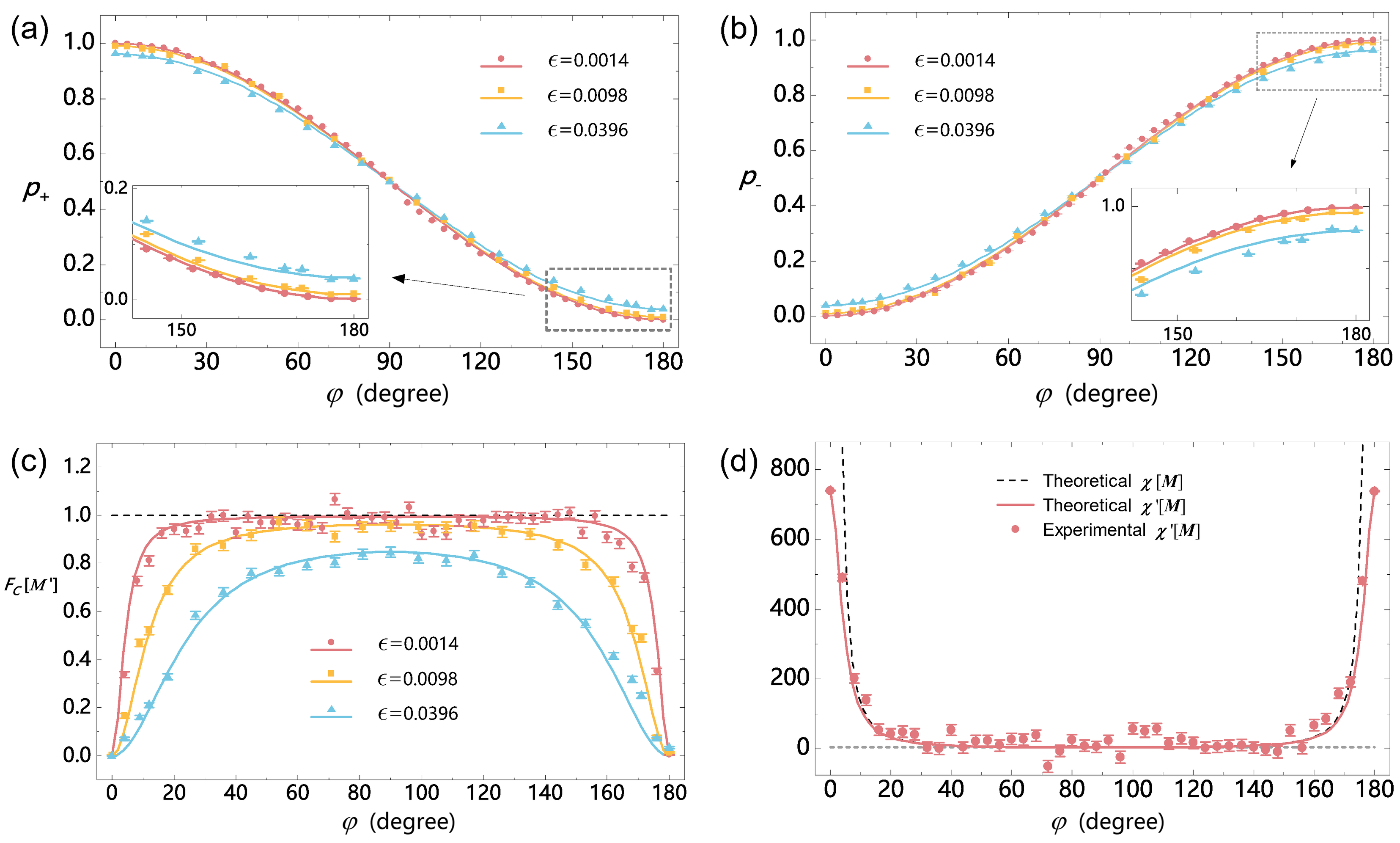}
  \caption{Estimation precision and FI MENOS. (a) The probability $p_+$ for varying measurements under different intensity of noise. (b) The probability $p_-$ for varying measurements under different intensity of noise. (c) The FI under different intensities of noise $\epsilon$, and the black dashed line is the QFI $F_Q[\bm{M}]=1$. Experimental results are close to the theoretical values. (d) The black dashed line is the theoretical FI MENOS $\chi[\bm{M}]$, the red dots (experiment) and line (theory) represent the $\chi'[\bm{M}]$ when $\epsilon=0.0014$. And the vertical coordinate value of dot gray line is 4. }\label{FI}
\end{figure*}

Parameter estimation is an asymptotically true theory, where CRB is only saturated when repetition factor $\nu$ is large. In our experiment, we conducted $\nu=1000\thicksim1200$ measurements to obtain the probabilities $p_\pm(\varphi)=\langle\psi_\theta|\hat{M}^\prime_\pm(\varphi)|\psi_\theta\rangle$. The phase estimator $\hat{\theta}$ was then obtained using the maximum likelihood method based on the probabilities of the measurement outcomes. This entire process was repeated 1000 times to acquire statistical information about the estimator. Figure\;\ref{FI}(a) and (b) present the experimental results of the probabilities, the actual value of the phase we set is $\theta=0$. The experimental estimation results are shown in Figure\;\ref{FI}(c), each data point of FI is obtained from the rescaled variances of the estimation results. It should be noticed that the variances are rescaled by multiplying them with the number of measurements, enabling a comparison with the CRB. As $\varphi$ approaches $0$ and $\pi$, the disparity between experimental FI and ideal QFI increases, indicating that FI is more susceptible to noise in this scenario even though the noise is small. Comparing the results presented in Figure\;\ref{FI}(a), (b) and (c), we can see that the estimation precision exhibits significant fluctuations across varying levels of noise intensity, despite minor alterations in probability outcomes. Notably, the Mach-Zehnder interferometer exhibits high sensitivity to environmental influences such as air flow, ambient temperature fluctuations, and experimental setup vibrations, all of which could significantly impact estimation results. These environmental factors are undesirable and uncontrollable, leading to deviations between experimental and theoretical results. Consequently, it is crucial to maintain a stable experimental environment.

Furthermore, we also plot the experimental results of FI MENOS in Figure\;\ref{FI}(d). Comparing the results in Figure\;\ref{FI}(c) and (d), the estimation results are consistent with the analysis of FI MENOS $\chi[\bm{M}]$, specifically, the region of increasing $\chi[\bm{M}]$ corresponds to a decrease in FI, and when $\varphi=\pi/2$, FI reaches its maximum while $\chi[\bm{M}]$ reaches its minimum. We could identify the robust optimal measurement is $\hat{M}_\pm(\pi/2)$. And three sets of experimental results ($\epsilon=0.0014,0.0098,0.0396$) show that the intensity of noise directly influences the decrease of FI.

Note that according to the definition of FI MENOS\;(\ref{FIMENOS}), the experimental noise $\bm{N}$ is required to maximize $\chi[\bm{M},\bm{N}]$, which is challenging in Mach-Zehnder interferometer, because the form of noise $\bm{N}$ in Mach-Zehnder interferometer is immutable. Fortunately, the noise we realized happens to maximize $\chi[\bm{M},\bm{N}]$, the FI MENOS in our experiment can be written as \cite{noise3}
\begin{eqnarray}
\chi[\bm{M}]=\chi[\hat{M}_\pm(\varphi),\hat{M}_\mp(\varphi)]=1+\cos^{-2}(\frac{\varphi}{2})+\tan^{-2}(\frac{\varphi}{2}).\nonumber\\
\end{eqnarray}
Then, it is sufficient to measure $\chi[\hat{M}_\pm(\varphi),\hat{M}_\mp(\varphi)]$. However, the intensity of noise is required to approach $0$ according to eq.\;(\ref{chiMN}), it means that the interference visibility $\mathcal{V}$ need to reach $1$. Achieving a visibility of 1 is nearly unattainable in actual experimental, therefore, our focus is on maximizing the interference visibility to the best extent possible. The quantity we have actually measured is
\begin{eqnarray}
\chi'[\bm{M}]=\frac{F_C[\hat{M}_\pm]-F_C[(1-\epsilon)\hat{M}_\pm+\epsilon\hat{M}_\mp]}{\epsilon F_C[\hat{M}_\pm]},\label{expFIMENOS}
\end{eqnarray}
where $\epsilon$ is quite small, $F_C[\hat{M}_\pm]$ is the theoretical FI corresponds to the perfect optimal measurement, and $F_C[(1-\epsilon)\hat{M}_\pm+\epsilon\hat{M}_\mp]$ is the actual FI we measured experimentally. Substituting the experimental FI and the intensity of noise $\epsilon=0.0014$ into eq.\;(\ref{expFIMENOS}), we obtain the experimental results of $\chi'[\bm{M}]$. As shown in Figure\;\ref{FI}(d), $\chi'[\bm{M}]$ closely approximates the theoretical $\chi[\bm{M}]$, particularly when $\varphi$ approaches $\pi/2$. The results match well with the theoretical model.

\begin{figure*}[t]
  \centering
  \includegraphics[width=1\textwidth]{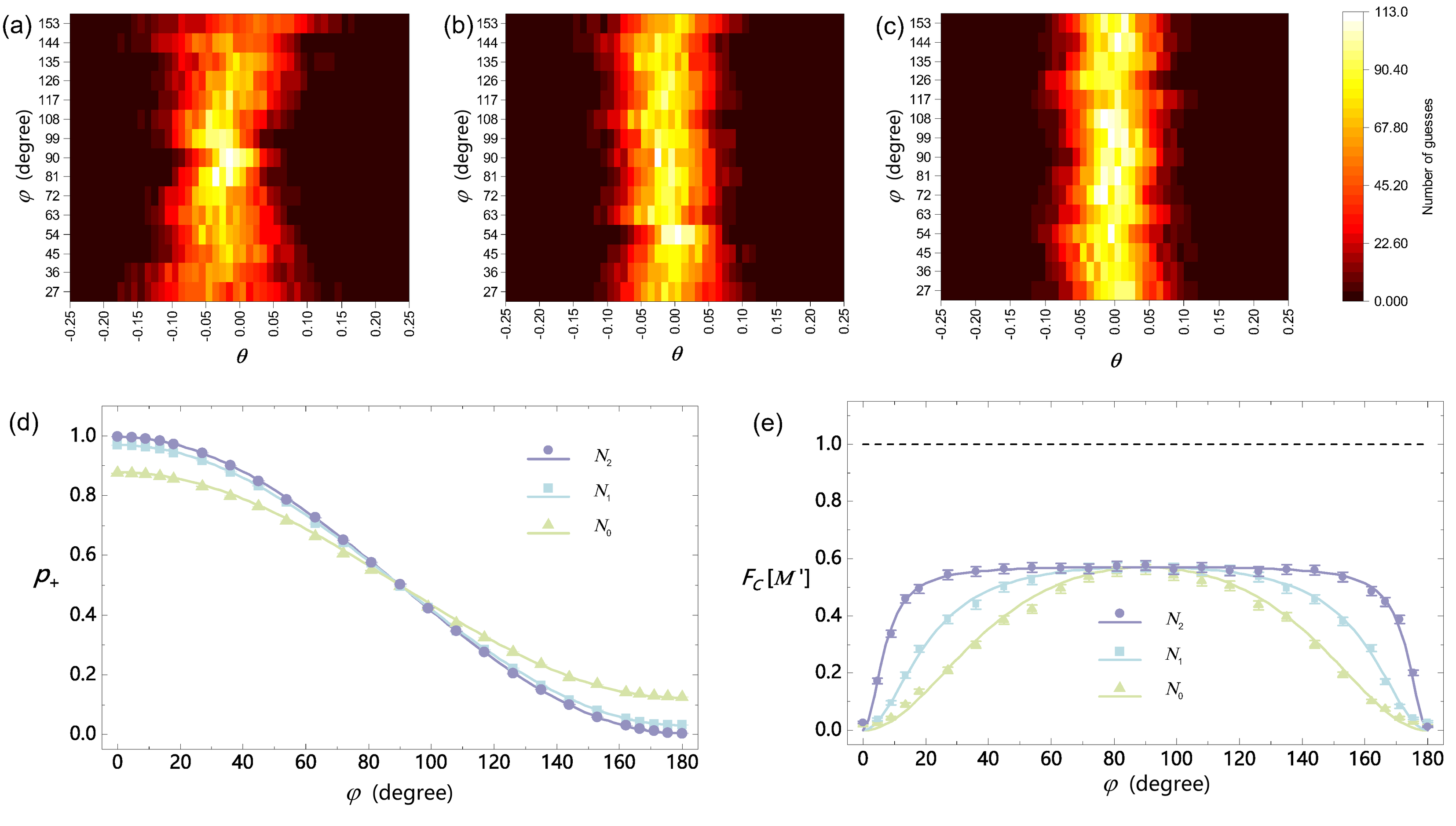}
  \caption{Estimation precision and FI MENOS. (a) The distribution of estimation results when the noise is $\bm{N}_0=\hat{U}_{H}(0^\circ)\hat{M}_\pm(\varphi)\hat{U}_{H}^\dagger(0^\circ)$. (b) The distribution of estimation results when the noise is $\bm{N}_1=\hat{U}_{H}(30^\circ)\hat{M}_\pm(\varphi)\hat{U}_{H}^\dagger(30^\circ)$. (c) The distribution of estimation results when the noise is $\bm{N}_2=\hat{U}_{H}(40^\circ)\hat{M}_\pm(\varphi)\hat{U}_{H}^\dagger(40^\circ)$. (d) The probability $p_+$ for varying measurements under different forms of noise. (e) The FI under different intensities of noise $\epsilon$, and the black dashed line is the QFI $F_Q[\bm{M}]=1$. Experimental results are close to the theoretical values.}\label{WPFI}
\end{figure*}

\section{Phase estimation under different forms of noise}\label{sec:6}
In this scenario, the probe state $|\psi_0\rangle$, parameter $\theta=0$ and evolution operator $\hat{U}_\theta$ remain the same as those set in the Mach-Zehnder interferometer experiment, but the imperfect measurement $\bm{\tilde{M}}$ does not have to be $(1-\epsilon)\hat{M}_\pm+\epsilon\hat{M}_\mp$. The exact form of noise $\epsilon\bm{N}$ is determined by the combination of HWPs and QWPs in transmitted path $a$, as shown in the bottom layer of Figure\;\ref{exp}(a). Here, we set $\bm{N}=\{\hat{U}_{H}(\alpha_{5})\hat{M}_\pm(\varphi)\hat{U}_{H}^\dagger(\alpha_{5})$\}. Therefore, we can introduce various forms of noise by changing the operator $\hat{U}_{H}(\alpha_{5})$, where $\hat{U}_{H}(\alpha_{5})$ is the Jones matrix of H5 and changed by rotating the angle $\alpha_{5}$ of H5. The rest of four wave plates in path $a$ and PBS4 constitute $\hat{M}_\pm(\varphi)$. Meanwhile, the combination of HWPs, QWPs and PBS4 in reflected path $b$ forms the component of ideal measurement $(1-\epsilon)\hat{M}_\pm(\varphi)$. The exact imperfect measurement $\bm{\tilde{M}}=\{\hat{M}'_\pm(\varphi)\}$ we realized in this experiment is $\hat{M}'_\pm(\varphi)=(1-\epsilon)\hat{M}_\pm(\varphi)+\epsilon\hat{U}_{H}(\alpha_{5})\hat{M}_\pm(\varphi)\hat{U}_{H}^\dagger(\alpha_{5})$. The intensity of noise $\epsilon$ is determined by the split ratio $r$ (R:T) of NPBS1, we have $\epsilon=1/(r+1)$. In our experiment, the exact values of split ratio and the intensity of noise are $r=7.1633$ and $\epsilon=0.1225$.

Figure 4(a), (b), and (c) depict the distribution of estimation results for three distinct forms of noise, which exhibit a higher concentration around $\varphi=90^\circ$. Owing to uncontrollable experimental errors (e.g., inaccuracy in wave plate angle), the centers of these distributions exhibit slight deviations from the experimentally preset values. Such slight deviation is acceptable, as the variance of the distribution is close to the theoretical prediction. Among them, the performance of estimation precision is worst for noise $\bm{N}_0$. Similarly, the estimation results are obtained via maximum likelihood method based on the measurement probabilities, the probabilities are presented in Figure\;\ref{WPFI}(d). We also plot the experimental precisions under different forms of noise in Figure\;\ref{WPFI}(e). Notably, the estimation precision exhibits the highest sensitivity to noise $\bm{N}_0$, which actually corresponds to the case $\chi[\bm{M}]$. The estimation precision under the other two types of noise does not deteriorate beyond the scenario characterized by $\chi[\bm{M}]$ for all $\varphi$, consistent with the predictions of $\chi[\bm{M},\bm{N}]$ for different forms of noise $\bm{N}$.
\begin{eqnarray}
\chi[\bm{M},\bm{N}_0]\leq\chi[\bm{M},\bm{N}_1]\leq\chi[\bm{M},\bm{N}_2]
\end{eqnarray}
Therefore, the advantage of the analysis based on FI MENOS, is that we do not need to consider particular form of noise, and the worst-case scenario has been taken into account.

\section{Discussions and conclusion}\label{sec:7}
It is worth mentioning that within the framework of FI MENOS theory, the parameter can still be estimated accurately in the presence of noise, i.e., $E[\hat{\theta}]=\theta$. However, it requires prior knowledge of both the intensity $\epsilon$ and form $\bm{N}$ of noise, such that we can eliminate the impact of noise when finding the estimator, although at the expense of reducing the QFI. In practical experiments and applications, the limited information available about the noise will lead to deviations between the estimated value and the true value of parameter \cite{biasestimator}, while also narrowing the dynamical range. We envisage that the approach of FI MENOS may be naturally extended to the case that the information of noise is unknown to investigate these metrics \cite{dynamic-range}. Moreover, recent research has demonstrated that the non-phase-covariant noise during the evolution process holds significant potential for enhancing estimation precision \cite{covariant-noise}. In general, noise holds significant research value in quantum metrology and merits further investigation.

In conclusion, we have experimentally implemented the phase estimation in polarizing Mach-Zehnder interferometer to study robust optimal measurements based on the theory of FI MENOS. The estimation precision across different measurements is consistent with the prediction of FI MENOS. The lager FI MENOS means that the corresponding measurement is more susceptible to noise. The impact of different forms noise on estimation precision is also considered, the experimental estimation precision still matches FI MENOS theory well. FI MENOS represents the worst-case scenario, allowing us not to consider a particular noise model. Based on the analysis of the FI MENOS, we could identify the most robust measurement among the various optimal measurements. Our experimental results are universal, as we discussed above, we can always analyze FI MENOS on Bloch sphere in the case of pure state model. This opens the door for establishing a robust scheme of quantum estimation.

%%%%%%%%%%%%%%%%%%%%%%%%%%%%%%%%%%%%%%%%%%%%%%%%%%%%%%%
%%% Acknowledgements. ??¡ìY
%%%%%%%%%%%%%%%%%%%%%%%%%%%%%%%%%%%%%%%%%%%%%%%%%%%%%%%
\section{Acknowledgements}
This work is supported by the Innovation Program for Quantum Science and Technology (Grant No.~2021ZD0301200), the National Natural Science Foundation of China (Grants No.~62475127 and No.~11821404), the Zhejiang Provincial Natural Science Foundation of China (Grant No. LZ25A040006), the  K.C. Wong Magna Fund in Ningbo University. S.K. acknowledges support from the National Science Center (Poland) (Grant No.2020/37/B/ST2/02134).\\

%%%%%%%%%%%%%%%%%%%%%%%%%%%%%%%%%%%%%%%%%%%%%%%%%%%%%%%
%%% Appendix sections. ??????, ????
%%%%%%%%%%%%%%%%%%%%%%%%%%%%%%%%%%%%%%%%%%%%%%%%%%%%%%%
%\section{Name}

%\end{appendix}

%\begin{appendices}
%\section{Appendix}
%\end{appendices}
%\appendix

%\appendix


\begin{thebibliography}{99}

\bibitem{computing1}H. Zhou, C. Zhao, M. Cain, D. Bluvstein, N. Maskara, C. Duckering, H.-Y. Hu, S.-T. Wang, A. Kubica, and M. D. Lukin, \href{https://doi.org/10.1038/s41586-025-09543-5}{Nature \textbf{646}, 303 (2025).}

\bibitem{computing2} H. Aghaee Rad, T. Ainsworth, R.N. Alexander, B. Altieri, M. F. Askarani, et al., \href{https://doi.org/10.1038/s41586-024-08406-9}{Nature \textbf{638}, 912 (2025).}

\bibitem{computing3}D.-Q. Ma, Q.-X. Jie, Y.-D. Hu, W.-Y. Zhu, Y.-C. Zhang, H.-J. Fan, X.-K. Zhong, G.-J. Chen, Y.-L. Zhang, T.-Y. Zhang, X.-F. Ren, L. Chen, Z.-B. Wang, G.-C. Guo, and C.-L. Zou, \href{https://doi.org/10.1007/s11433-025-2804-2}{Sci. China-Phys. Mech. Astron. \textbf{69}, 220314 (2026).}

\bibitem{wave1}S. Pedalino, B. E. Ram\'{\i}rez-Galindo, R. Ferstl, K. Hornberger, M. Arndt, and S. Gerlich, \href{https://doi.org/10.1038/s41586-025-09917-9}{Nature \textbf{649}, 866 (2026).}

\bibitem{wave2}X. Yang, X. Yu, L. Li, X. Zhao, T. Zheng, C. Zhang, C.-F. Li, and G.-C. Guo, \href{https://doi.org/10.1007/s11433-024-2587-y}{Sci. China-Phys. Mech. Astron. \textbf{68}, 250312 (2025)}

\bibitem{communication1}Y. Zheng, H. Wang, X. Jia, J. Huang, H. Yuan, C. Zhai, J. Dai, J. Shi, L. Zhang, X. Zhang, M. Zhuang, J. Liu, J. Mao, T. Dai, Z. Fu, Y. Jiao, Y. Shi, D. Dai, X. Wang, Y. Li, Q. Gong, Z. Yuan, L. Chang, and J. Wang,\href{https://doi.org/10.1038/s41586-026-10152-z}{Nature \textbf{651}, 68 (2026)}

\bibitem{communication2}M.-Y. Lv, X.-M. Hu, N.-F. Gong, T.-J. Wang, Y. Guo, B.-H. Liu, Y.-F. Huang, C.-F. Li, and G.-C. Guo, \href{https://doi.org/10.1007/s11433-023-2286-8}{Sci. China-Phys. Mech. Astron. \textbf{67}, 230311 (2024)}
\bibitem{atmm}D. S. Weiss, B. C. Young, and S. Chu, \href{https://doi.org/10.1103/PhysRevLett.70.2706}{Phys. Rev. Lett. \textbf{70}, 2706 (1993).}

\bibitem{fstruc1}R. Bouchendira, P. Clad\'{e}, S. Guellati-Kh\'{e}lifa, F. Nez, and F. Biraben, \href{https://doi.org/10.1103/PhysRevLett.106.080801}{Phys. Rev. Lett. \textbf{106}, 080801 (2011).}

\bibitem{fstruc2}R.~H. Parker, C.~Yu, W.~Zhong, B.~Estey, and H.~M\"{u}ller, \href{https://www.science.org/doi/abs/10.1126/science.aap7706}{Science \textbf{360}, 191 (2018).}

\bibitem{dakm1}B.~Elder, J.~Khoury, P.~Haslinger, M.~Jaffe, H.~M\"{u}ller, and P.~Hamilton, \href{https://link.aps.org/doi/10.1103/PhysRevD.94.044051}{Phys. Rev. D \textbf{94}, 044051 (2016).}

\bibitem{dakm2}P.~Hamilton, M.~Jaffe, P.~Haslinger, Q.~Simmons, H.~M\"{u}ller, and J.~Khoury, \href{https://www.science.org/doi/abs/10.1126/science.aaa8883}{Science \textbf{349}, 849 (2015).}

\bibitem{gw1}S.~Dimopoulos, P.~W. Graham, J.~M. Hogan, M.~A. Kasevich, and S.~Rajendran, \href{https://link.aps.org/doi/10.1103/PhysRevD.78.122002}{Phys. Rev. D \textbf{78}, 122002 (2008).}

\bibitem{gw2}B.~P. Abbott, R.~Abbott, T.~D. Abbott, M.~R. Abernathy, F.~Acernese, K.~Ackley, C.~Adams, T.~Adams, P.~Addesso, R.~X. Adhikari, \emph{et~al.}, \href{https://link.aps.org/doi/10.1103/PhysRevLett.116.061102}{Phys. Rev. Lett. \textbf{116}, 061102 (2016).}

\bibitem{gw3}J.~M. Hogan and M.~A. Kasevich, \href{https://link.aps.org/doi/10.1103/PhysRevA.94.033632}{Phys. Rev. A \textbf{94}, 033632 (2016).}

\bibitem{gw4}D.~Ganapathy, W.~Jia, M.~Nakano, V.~Xu, N.~Aritomi, T.~Cullen, N.~Kijbunchoo, S.~E. Dwyer, A.~Mullavey \emph{et~al.}, \href{https://doi.org/10.1103/PhysRevX.13.041021}{Phys. Rev. X \textbf{13}, 041021 (2023).}


\bibitem{qem1}C.~W. Helstrom, \emph{Quantum detection and estimation theory} (Academic Press, 1976).

\bibitem{qem2}A.~S. Holevo, \emph{Probabilistic and statistical aspects of quantum theory} (North-Holland Publishing Company, 1982).

\bibitem{qem3}S.~L. Braunstein and C.~M. Caves, \href{https://link.aps.org/doi/10.1103/PhysRevLett.72.3439}{Phys. Rev. Lett. \textbf{72}, 3439 (1994).}

\bibitem{qem4}S.~L. Braunstein, C.~M. Caves, and G.~Milburn, \href{https://www.sciencedirect.com/science/article/pii/S0003491696900408}{Ann. Phys \textbf{247}, 135 (1996).}

\bibitem{HL1}J.~J. Bollinger, W.~M. Itano, D.~J. Wineland, and D.~J. Heinzen, \href{https://link.aps.org/doi/10.1103/PhysRevA.54.R4649}{Phys. Rev. A \textbf{54}, R4649 (1996).}

\bibitem{HL2}V.~Giovannetti, S.~Lloyd, and L.~Maccone, \href{https://www.science.org/doi/abs/10.1126/science.1104149}{Science \textbf{306}, 1330 (2004).}

\bibitem{HL3}V.~Giovannetti, S.~Lloyd, and L.~Maccone, \href{https://link.aps.org/doi/10.1103/PhysRevLett.96.010401}{Phys. Rev. Lett. \textbf{96}, 010401 (2006).}

\bibitem{HL4}Q.~Liu, Z.~Hu, H.~Yuan, and Y.~Yang, \href{https://link.aps.org/doi/10.1103/PhysRevLett.130.070803}{Phys. Rev. Lett. \textbf{130}, 070803 (2023).}

\bibitem{entgHL1}T.~Nagata, R.~Okamoto, J.~L. O'Brien, K.~Sasaki, and S.~Takeuchi, \href{https://www.science.org/doi/abs/10.1126/science.1138007}{Science \textbf{316}, 726 (2007).}

\bibitem{entgHL2}G.-Y. Xiang, B.~L. Higgins, D.~Berry, H.~M. Wiseman, and G.~Pryde, \href{https://doi.org/10.1038/nphoton.2010.268}{Nature Photon \textbf{5}, 43(2011).}

\bibitem{seqHL1}B.~L. Higgins, D.~W. Berry, S.~D. Bartlett, H.~M. Wiseman, and G.~J. Pryde, \href{https://doi.org/10.1038/nature06257}{Nature \textbf{450}, 393 (2007).}

\bibitem{seqHL2}H. Yuan, C.-H. F. Fung, \href{https://doi.org/10.1038/nature06257}{Phys. Rev. Lett. \textbf{115}, 110401 (2015).}

\bibitem{seqHL3}D. Braun, G. Adesso, F. Benatti, R. Floreanini, U. Marzolino, M. W. Mitchell, and S. Pirandola, \href{https://doi.org/10.1103/RevModPhys.90.035006}{Rev. Mod. Phys. \textbf{90}, 035006 (2018).}

\bibitem{seqHL4}Z. Hou, R.-J. Wang, J.-F. Tang, H. Yuan, G.-Y. Xiang, C.-F. Li, and G.-C. Guo, \href{https://link.aps.org/doi/10.1103/PhysRevLett.123.040501}{Phys. Rev. Lett. \textbf{123}, 040501 (2019).}

\bibitem{seqHL5}Z. Hou, J.-F. Tang, H. Chen, H. Yuan, G.-Y. Xiang, C.-F. Li  and G.-C. Guo, \href{https://www.science.org/doi/abs/10.1126/sciadv.abd2986}{Sci. Adv. \textbf{7}, eabd2986 (2021).}

\bibitem{supHL1}X. Zhao, Y. Yang, and G. Chiribella, \href{https://doi.org/10.1103/PhysRevLett.124.190503}{Phys. Rev. Lett. \textbf{124}, 190503 (2020).}

\bibitem{supHL2}P. Yin, X. Zhao, Y. Yang, Y. Guo, W.-H. Zhang, G.-C. Li, Y.-J. Han, B.-H. Liu, J.-S. Xu, G. Chiribella, G. Chen, C.-F. Li and G.-C. Guo, \href{https://doi.org/10.1038/s41567-023-02046-y}{Nat. Phys. \textbf{19}, 1122 (2023).}

\bibitem{supHL3}S. Kurdzia{\l}ek, W. G\'{o}recki, F. Albarelli, and R. Demkowicz-Dobrza\'{n}ski, \href{https://doi.org/10.1103/PhysRevLett.131.090801}{Phys. Rev. Lett. \textbf{131}, 090801 (2023).}

\bibitem{nqs1}J. Wiersig, \href{https://doi.org/10.1103/PhysRevLett.112.203901}{Phys. Rev. Lett. \textbf{112}, 203901 (2014).}

\bibitem{nqs2}J. Wiersig, \href{https://doi.org/10.1103/PhysRevA.93.033809}{Phys. Rev. A \textbf{93}, 033809 (2016).}

\bibitem{nqs3}Z.-P. Liu1, J. Zhang, \c{S}. K. \"{O}zdemir, B. Peng, H. Jing, X.-Y. L\"{u}, C.-W. Li, L. Yang, F. Nori, \href{https://doi.org/10.1103/PhysRevLett.117.110802}{Phys. Rev. Lett. \textbf{117}, 110802 (2016).}

\bibitem{nqs4}W. Chen, \c{S}. K. \"{O}zdemir, G. Zhao, J. Wiersig and L. Yang, \href{https://doi.org/10.1038/nature23281}{Nature \textbf{548}, 192 (2017).}

\bibitem{nqs5}H. Hodaei, A. U. Hassan, S. Wittek, H. Garcia-Gracia, R. El-Ganainy, D. N. Christodoulides and M. Khajavikhan, \href{https://doi.org/10.1038/nature23280}{Nature \textbf{548}, 187 (2017).}

\bibitem{nqs6}X. Yu and C. Zhang, \href{https://doi.org/10.1103/PhysRevA.108.022215}{Phys. Rev. A \textbf{108}, 022215 (2023).}

\bibitem{nhqsXinglei}X. Yu, X. Zhao, L. Li, X.-M. Hu, X. Duan, H. Yuan, and C. Zhang, \href{https://www.science.org/doi/10.1126/sciadv.adk7616}{Science Advances \textbf{10}, eadk7616 (2024).}

\bibitem{noise1}A. Datta, L. Zhang, N. Thomas-Peter, U. Dorner, B. J. Smith, and I. A. Walmsley, \href{https://doi.org/10.1103/PhysRevA.83.063836}{Phys. Rev. A \textbf{83}, 063836 (2011).}

\bibitem{sensor}C. L. Degen, F. Reinhard, and P. Cappellaro, \href{https://doi.org/10.1103/RevModPhys.89.035002}{Rev. Mod. Phys. \textbf{89}, 035002 (2017).}

\bibitem{noise2}Y. L. Len, T. Gefen, A. Retzker and J. Ko{\l}ody\'{n}ski, \href{https://doi.org/10.1038/s41467-022-33563-8}{Nat Commun \textbf{13}, 6971 (2022).}

\bibitem{noise3}S. Kurdzia{\l}ek and Rafa{\l} Demkowicz-Dobrza\'{n}ski, \href{https://doi.org/10.1103/PhysRevLett.130.160802}{Phys. Rev. Lett. \textbf{130}, 160802 (2023).}

\bibitem{an1}K. Bai, Z. Peng, H.-G. Luo, and J.-H. An, \href{https://doi.org/10.1103/PhysRevLett.123.040402}{Phys. Rev. Lett. \textbf{123}, 040402 (2019).}

\bibitem{an2}S.-Y. Bai and J.-H. An, \href{https://doi-org.easyaccess1.lib.cuhk.edu.hk/10.1103/PhysRevLett.131.050801}{Phys. Rev. Lett. \textbf{131}, 050801 (2023).}

\bibitem{qcom1}D. Leibfried, B. DeMarco, V. Meyer, D. Lucas, M. Barrett, J. Britton, W. M. Itano, B. Jelenkovi\'{c}, C. Langer, T. Rosenband and D. J. Wineland, \href{https://doi.org/10.1038/nature01492}{Nature \textbf{422}, 412 (2003).}

\bibitem{qcom2}M. Urbanek, B. Nachman, V. R. Pascuzzi, A. He, C. W. Bauer, and W. A. de Jong, \href{https://doi.org/10.1103/PhysRevLett.127.270502}{Phys. Rev. Lett. \textbf{127}, 270502 (2021).}

\bibitem{qcom3}R. Harper, S. T. Flammia and J. J. Wallman, \href{https://doi.org/10.1038/s41567-020-0992-8}{Nat. Phys. \textbf{16}, 1184 (2020).}

\bibitem{qcom4}Y. Kim, A. Eddins, S. Anand, K. X. Wei, E. v. d. Berg, S. Rosenblatt, H. Nayfeh, Y. Wu, M. Zaletel, K. Temme and A. Kandala, \href{https://doi.org/10.1038/s41586-023-06096-3}{Nature \textbf{618}, 500 (2023).}

\bibitem{qcommu1}X.-M. Hu, C. Zhang, Y. Guo, F.-X. Wang, W.-B. Xing, C.-X. Huang, B.-H. Liu, Y.-F. Huang, C.-F. Li \emph{et~al.}, \href{https://doi.org/10.1103/PhysRevLett.127.110505}{Phys. Rev. Lett. \textbf{127}, 110505 (2021).}

\bibitem{qcommu2}F. Bouchard, D. England, P. J. Bustard, K. L. Fenwick, E. Karimi, K. Heshami, and B. Sussman, \href{https://doi.org/10.1103/PhysRevApplied.15.024027}{Phys. Rev. Applied \textbf{15}, 024027 (2021).}

\bibitem{qcommu3}L.-P. Lamoureux, E. Brainis, N. J. Cerf, Ph. Emplit, M. Haelterman, and S. Massar, \href{https://doi.org/10.1103/PhysRevLett.94.230501}{Phys. Rev. Lett. 94, 230501 (2025).}

\bibitem{qcommu4}K. B. A. Dragan, W. Wasilewski, and C. Radzewicz, \href{https://doi.org/10.1103/PhysRevLett.92.257901}{Phys. Rev. Lett. \textbf{92}, 257901 (2004).}

\bibitem{qcommu5}G. Rubino, L. A. Rozema, D. Ebler, H. Kristj\'{a}nsson, S. Salek, P. A. Gu\'{e}rin, A. A. Abbott, C. Branciard, \v{C}. Brukner, \href{https://doi.org/10.1103/PhysRevResearch.3.013093}{Phys. Rev. Research \textbf{3}, 013093 (2021).}

\bibitem{qcontr1}A. Soare, H. Ball, D. Hayes, J. Sastrawan, M. C. Jarratt, J. J. McLoughlin, X. Zhen, T. J. Green and M. J. Biercuk, \href{https://doi.org/10.1038/nphys3115}{Nature Phys \textbf{10}, 825 (2014).}

\bibitem{qcontr2}M. Rossi, D. Mason, J. Chen, Y. Tsaturyan and A. Schliesser, \href{https://doi.org/10.1038/s41586-018-0643-8}{Nature \textbf{563}, 53 (2018).}

\bibitem{qcontr3}L. Magrini, P. Rosenzweig, C. Bach, A. Deutschmann-Olek, S. G. Hofer, S. Hong, N. Kiesel, A. Kugi and M. Aspelmeyer, \href{https://doi.org/10.1038/s41586-021-03602-3}{Nature \textbf{595}, 373 (2021).}

\bibitem{qcontr4}Z. Zhou, R. Sitler, Y. Oda, K. Schultz, and G. Quiroz, \href{https://doi.org/10.1103/PhysRevLett.131.210802}{Phys. Rev. Lett. \textbf{131}, 210802 (2023).}

\bibitem{noiseqm1}K. Wang, X. Wang, X. Zhan, Z. Bian, J. Li, B. C. Sanders, and P. Xue, \href{https://doi.org/10.1103/PhysRevA.97.042112}{Phys. Rev. A \textbf{97}, 042112 (2018).}

\bibitem{noiseqm2}X. Long, W.-T. He, N.-N. Zhang, K. Tang, Z. Lin, H. Liu, X. Nie, G. Feng, J. Li \emph{et al.}, \href{https://doi.org/10.1103/PhysRevLett.129.070502}{Phys. Rev. Lett. \textbf{129}, 070502 (2022).}

\bibitem{multi-FIMENOS}F. Albarelli, I. Gianani, M. G. Genoni and M. Barbieri, \href{https://doi.org/10.1103/PhysRevA.110.032436}{Phys. Rev. A \textbf{110}, 032436 (2024).}

\bibitem{optm4}S. Zhou, C.-L. Zou and L. Jiang, \href{https://iopscience.iop.org/article/10.1088/2058-9565/ab71f8/meta}{Quantum Sci. Technol. \textbf{5} 025005 (2020).}

\bibitem{sps}T. Kim, M. Fiorentino, and F. N. C. Wong, \href{https://doi.org/10.1103/PhysRevA.73.012316}{Phys. Rev. A \textbf{73}, 012316 (2006).}

\bibitem{biasestimator}K. Yamamoto, S. Endo, H. Hakoshima, Y. Matsuzaki, and Y. Tokunaga, \href{https://doi.org/10.1103/PhysRevLett.129.250503}{Phys. Rev. Lett. \textbf{129}, 250503 (2022).}

\bibitem{dynamic-range}Q. Liu, M. Xue, M. Radzihovsky, X. Li, D. V. Vasilyev, L.-N. Wu, and V. Vuleti\'{c},\href{https://doi-org.easyaccess1.lib.cuhk.edu.hk/10.1103/25ds-9724}{Phys. Rev. Lett. \textbf{135}, 040801 (2024).}

\bibitem{covariant-noise}J.-X. Peng, B. Zhu, W. Zhang and K. Zhang,\href{https://doi.org/10.1103/PhysRevLett.133.090801}{Phys. Rev. Lett. \textbf{133}, 090801 (2024).}

\end{thebibliography}
\end{document}